\documentclass[12pt,english]{article}
\usepackage[T1]{fontenc}
\usepackage[latin9]{inputenc}
\usepackage{geometry}
\usepackage{babel}
\usepackage{amsmath}
\usepackage{amssymb}
\usepackage{setspace}
\PassOptionsToPackage{normalem}{ulem}
\usepackage{ulem}
\usepackage[unicode=true,pdfusetitle,
 bookmarks=true,bookmarksnumbered=false,bookmarksopen=false,
 breaklinks=false,pdfborder={0 0 1},backref=false,colorlinks=false]
 {hyperref}

\makeatletter
\newcommand{\lyxaddress}[1]{
	\par {\raggedright #1
	\vspace{1.4em}
	\noindent\par}
}

\makeatother

\begin{document}
\title{$\mathbb{Z}_{2}-$Graded Lie Algebra of Quaternions and Superconformal
Algebra in $D=4$ dimensions }
\author{Bhupendra C. S. Chauhan$^{1}$, Pawan Kumar Joshi$^{2}$, B.C. Chanyal$^{3}$}
\maketitle

\lyxaddress{\begin{center}
$^{1,2}$Department of Physics, Kumaun University, S. S. J. Campus,
Almora \textendash{} 263601 (Uttarakhand), India.\\
$^{3}$Department of Physics, G.B. Pant University of Agriculture
and Technology, Pantnagar-263145 (Uttarakhand ), India.\\
 Email: $^{1}$bupendra.123@gmail.com \\
$^{2}$pj4234@gmail.com\\
$^{3}$bcchanyal@gmail.com
\par\end{center}}
\begin{abstract}
In the present discussion, we have studied the $\mathbb{Z}_{2}-$$grading$
of quaternion algebra $(\mathbb{H})$. We have made an attempt to
extend the quaternion Lie algebra to the graded Lie algebra by using
the matrix representations of quaternion units. The generalized Jacobi
identities of $\mathbb{Z}_{2}-graded$ algebra then result in symmetric
graded partners $(N_{1},N_{2},N_{3})$. The graded partner algebra
$(\mathcal{F})$ of quaternions $(\mathbb{H})$ thus has been constructed
from this complete set of graded partner units $(N_{1},N_{2},N_{3})$,
and $N_{0}=C$. Keeping in view the algebraic properties of the graded
partner algebra $(\mathcal{F})$, the $\mathbb{Z}_{2}-graded$ superspace
$(S^{l,m})$ of quaternion algebra $(\mathbb{H})$ has been constructed.
It has been shown that the antiunitary quaternionic supergroup $UU_{a}(l;m;\mathbb{H})$
describes the isometries of $\mathbb{Z}_{2}-graded$ superspace $(S^{l,m})$.
The Superconformal algebra in $D=4$ dimensions is then established,
where the bosonic sector of the Superconformal algebra has been constructed
from the quaternion algebra $(\mathbb{H})$ and the fermionic sector
from the graded partner algebra $(\mathcal{F})$.
\end{abstract}

\section{Introduction:}

In order to unify the symmetry of the Poincaré group with some internal
groups, several attempts have been made. Coleman and Mandula \cite{key-1}
in $1967$ set a restriction on the incorporation of Poincaré symmetry
into the internal symmetry group, up to which this unification is
possible. But actually, this doesn't offer any unification of the
Poincaré group with the internal symmetry group. This theorem, which
is called the no-go theorem of Coleman and Mandula, is based on the
extension of the symmetry of the $S$-matrix under the assumption
of physical conditions of locality, causality, positive energy, and
for finite numbers of particles. Weiss and Zumino \cite{key-2,key-3}
realized that the unification of the Poincaré group with the internal
symmetry group is possible by introducing anti-commutation relations
of supersymmetric charges into the theory, which relate the fermions
to bosons. However, its proof has been established by Haag, Lapuszanski,
and Sohnius \cite{key-3,key-4}. 

Thus, supersymmetric field theories arise as the maximum symmetry
of the $S$-matrix that is possible. This is the largest extension
of the Lie algebra of the Poincaré group and the internal symmetry
group, which has not only commutators but also anti-commutators of
supercharges that generate the supersymmetric transformations \cite{key-3,key-4}.
Since this theory can be a possible answer for most of the hierarchy
problems \cite{key-4} of the standard model, the unification of gravitation,
dark matter, and dark energy, several attempts have been made experimentally
in search of this symmetry. But it has not been confirmed yet. However,
it has always been an interesting theory searched by the theoretical
physicists in an attempt to unify the fundamental forces of nature. 

In the theoretical literature on supersymmetric field theories, L.
Brink et al. \cite{key-5} established the fact that supersymmetry
is only possible for the cases of dimensions $D=2,4,6,$ and $10,$
called critical dimensions. They showed that the action is supersymmetric
only for $D=4,6,$ and $10$ dimensions without the inclusion of further
fields. Also, it has been shown that \cite{key-3,key-4,key-5} non-abelian
Yang\textendash Mills fields with minimal coupling to mass-less spinors
are supersymmetric if and only if the dimension of space-time is $3,4,6$,
and $10$. 

On the other hand, according to the celebrated Hurwitz theorem \cite{key-6}-\cite{key-11},
there are four normed-division algebras consisting of $\mathbb{R}$
(real numbers), $\mathbb{C}$ (complex numbers), $\mathbb{H}$ (quaternions)\cite{key-6}-\cite{key-14},
and $\mathbb{O}$ (octonions) \cite{key-10}-\cite{key-17}. All four
algebras are alternative with antisymmetric associators. Real numbers
and complex numbers are limited to only two dimensions; quaternions
are extended to four dimensions; and octonions represent eight dimensions.
Keeping this in mind, many authors \cite{key-18}-\cite{key-30} tried
to establish a connection between the supersymmetric theories in critical
dimensions and the four normed-division algebras. One such connection
was studied by Kugo-Townsend \cite{key-19}, who established a relation
between the supersymmetric algebra in various dimensions and the four-division
algebras. This was further extended and generalized by Jerzy Lukierski
et al. \cite{key-20,key-21} and other authors \cite{key-22}-\cite{key-30}
as well. It is summarized that the algebras $\mathbb{R}$,$\mathbb{C}$,$\mathbb{H}$,$\mathbb{O}$
can be useful for the description of supersymmetric field theories
in higher dimensions. 

Keeping in view the connection between the normed division algebras
($\mathbb{R},$$\mathbb{C}$,$\mathbb{H}$, $\mathbb{O}$) and supersymmetric
theories, in the present paper, we have made an attempt to study the
$\mathbb{Z}_{2}-grading$ \cite{key-31} of quaternion algebra using
matrix representations of quaternion basis units $(e_{1},e_{2},e_{3})$. 

The whole paper is arranged in seven sections, including the introduction.
Section $2$ contains a basic introduction to quaternion algebra and
the matrix representations of its basis units $(e_{1},e_{2},e_{3})$.
In Section 3, we have studied the graded Lie algebra of quaternions
$(\mathbb{H})$, where we have defined the graded partner matrices
evaluated by the grading of quaternion algebra $(\mathbb{H})$. In
Section $4$, we have studied the relations between quaternion algebra
$(\mathbb{H})$ and the proposed graded partner algebra $(\mathcal{F})$
composed by the graded partner matrices ($N_{1},N_{2},N_{3}$, and
$N_{0}=C$). Section $5$ contains the dynamics of the superspace
constructed from the quaternion algebra $(\mathbb{H})$ and the graded
partner algebra $(\mathcal{F})$ under the quaternionic supergroups
\cite{key-20}. In Section $6$, the Superconformal algebra in $D=4$
has been established in terms of quaternions $(\mathbb{H})$ and its
graded partner algebra $(\mathcal{F})$. Section $7$ is for discussion
and conclusions.

\section{Quaternion Algebra $(\mathbb{H})$ (Definition): }

According to the celebrated Hurwitz theorem \cite{key-7,key-11},
there exist four normed division algebras: $\mathbb{R}$,$\mathbb{C}$,$\mathbb{H}$,
and $\mathbb{O}$, respectively, named as the algebras of real numbers,
complex numbers, quaternions, and octonions. The quaternion algebra
$(\mathbb{H})$ is the second largest normed division algebra, which
is non-commutative but associative. Any element of this quaternion
algebra $(\mathbb{H})$ is called a quaternion, which is expressed
over the field of real numbers as
\begin{align}
Q= & q^{0}e_{0}+q^{1}e_{1}+q^{2}e_{2}+q^{3}e_{3}.\quad\quad\quad(Q\in\mathbb{H}).\label{eq:1}
\end{align}
Where $q^{0}$ and $q^{j}\,\:(\forall j=1,2,3)$ are the real numbers
and the quaternion basis elements $e_{1},e_{2},e_{3}$ satisfy the
following multiplication rules:

\begin{align}
e_{i}e_{j}= & -\delta_{ij}+\in_{ijk}e_{k}\;\;\;\;(\forall i,j,k=1\,to\,3).\label{eq:2}
\end{align}
The $\in_{ijk}$ is a Levi-Civita tensor, which is totally antisymmetric
and has a value of $+1$ for the permutations: $(ijk)=(123),(231),(312)$. 

The quaternion conjugate is defined by $\overline{Q}=q^{0}e_{0}-q^{1}e_{1}-q^{2}e_{2}-q^{3}e_{3}\:(q\in\mathbb{H}).$
It is to be noted that the quaternion conjugation operation satisfies
the following composition rule:
\begin{align}
\overline{(Q_{1}Q_{2})}= & \overline{Q}_{2}\overline{Q}_{1}\:\:\:\:\:\:\:\:\:\:(\forall Q_{1},Q_{2}\in\mathbb{H}).\label{eq:3}
\end{align}
The norm of a quaternion is positive-definite and computed as

\begin{align}
N(Q)= & Q\overline{Q}=\overline{Q}Q=\left|Q\right|=(q^{0})^{2}+(q^{1})^{2}+(q^{2})^{2}+(q^{3})^{2}\geq0.\label{eq:4}
\end{align}
Quaternion is the second-highest normed division algebra; the norm
of quaternion satisfies the following multiplication rule:
\begin{align}
N(Q_{1})N(Q_{2})= & N(Q_{1}Q_{2})\:\:\:\:\:\:\:\:\:\:(\forall Q_{1},Q_{2}\in\mathbb{H}).\label{eq:5}
\end{align}
Following the eq. (\ref{eq:4}), the inverse of a quaternion is defined
by $Q^{-1}=\frac{\overline{Q}}{\left|Q\right|}$. Meanwhile, the inverse
of the multiplication of two quaternions has the following property:
\begin{align}
(Q_{1}Q_{2})^{-1}= & Q_{1}^{-1}Q_{2}^{-1}\:\:\:\:\:\:\:\:\:\:(\forall Q_{1},Q_{2}\in\mathbb{H}).\label{eq:6}
\end{align}
A $2\times2$ quaternionic matrix is defined as
\begin{align}
H= & \left[\begin{array}{cc}
h_{1} & h_{2}\\
h_{3} & h_{4}
\end{array}\right]\:\:\:\:\:\:\:\:\:\:(\forall h_{1},h_{2},h_{3},h_{4}\in\mathbb{H}).\label{eq:7}
\end{align}
Where $h_{1},h_{2},h_{3},$ and $h_{4}$ are the elements of quaternion
algebra $(\mathbb{H}).$ The Hermitian conjugate of this quaternionic
matrix is defined as
\begin{align}
H^{\dagger}= & \left[\begin{array}{cc}
\overline{h}_{1} & \overline{h}_{3}\\
\overline{h}_{2} & \overline{h}_{4}
\end{array}\right],\label{eq:8}
\end{align}

where $\overline{h}_{1},\overline{h}_{2},\overline{h}_{3},\overline{h}_{4}$
are the quaternion conjugates of $h_{1},h_{2},h_{3},h_{4}$, respectively.
The Hermitian conjugation operation in quaternionic matrices satisfies
the following composition rule: 
\begin{align}
(H_{1}H_{2})^{\dagger}= & H_{2}^{\dagger}H_{1}^{\dagger}.\label{eq:9}
\end{align}
In a $4\times4$ real matrix representation, the quaternion basis
elements are defined \cite{key-20} as

\begin{align}
e_{1}= & \left[\begin{array}{cccc}
0 & 1 & 0 & 0\\
-1 & 0 & 0 & 0\\
0 & 0 & 0 & -1\\
0 & 0 & 1 & 0
\end{array}\right],\:\:e_{2}=\left[\begin{array}{cccc}
0 & 0 & 1 & 0\\
0 & 0 & 0 & 1\\
-1 & 0 & 0 & 0\\
0 & -1 & 0 & 0
\end{array}\right],\,\,e_{3}=\left[\begin{array}{cccc}
0 & 0 & 0 & 1\\
0 & 0 & -1 & 0\\
0 & 1 & 0 & 0\\
-1 & 0 & 0 & 0
\end{array}\right].\label{eq:10}
\end{align}
These basis elements satisfy the following commutation algebra:

\begin{align}
[e_{i},e_{j}]= & 2\in_{ijk}e_{k}\quad\quad\quad(\forall i,j,k=1\,to\,3).\nonumber \\{}
[e_{p},[e_{q},e_{r}]]+[e_{q},[e_{r},e_{p}]]+[e_{r},[e_{p},e_{q}]]= & 0\quad\quad\quad\quad\quad\quad\quad\:\:(\forall p,q,r=1\,to\,3).\label{eq:11}
\end{align}
The matrix representations of quaternion basis units defined in eq.
(\ref{eq:10}) are real and unitary, i.e., $e_{i}^{T}=-e_{i}=e_{i}^{-1}=e_{i}^{\dagger}$
$(\forall i=1,2,\,and\,3)$. 

\section{Graded Lie Algebra of Quaternions $(\mathbb{H})$:}

The $Z_{2}$-graded algebra $L$ is the direct sum of two algebras
\cite{key-30,key-31}, with $L_{0}\oplus L_{1}$ having the following
properties:

(i) $L_{0}$ is an even Lie algebra with degree $g(L_{0})=0$ and
$L_{0}\times L_{0}\rightarrow L_{0}$

(ii) $L_{1}$ is an odd Lie algebra with degree $g(L_{1})=1$ and
$L_{0}\times L_{1}\rightarrow L_{1}$ , $L_{1}\times L_{1}\rightarrow L_{0}$

(iii) Representation of $L_{0}$ in $dim\,L_{1}\times dim\,L_{1}$ 

Now to construct $Z_{2}$-graded algebra for quaternions, we take
the $4\times4$ dimensional representation of quaternions defined
in eq. (\ref{eq:10}). These representations of quaternions satisfy
the Lie algebra described in eq. (\ref{eq:11}). This Lie algebra
is closed, and hence it is an even algebra, $L_{0}$, that constructs
the bosonic part of the graded Lie algebra $(L)$. While the odd algebra
$L_{1}$ is not closed, as one can see in the second (ii) axiom of
the above-described properties of $Z_{2}-$graded algebra, In a mathematical
way, we now summarize these assertions for the $Z_{2}$-graded algebra
$(L)$ of quaternions $(\mathbb{H})$ as:

\begin{align}
1.\:L_{0}= & \{e_{i}\in L_{0},\;\;[e_{i},e_{j}]=2\in_{ijk}e_{k},\:(\forall i,j,k=1\,to\,3)\}\nonumber \\
2.\:L_{1}= & \{Q_{a}\in L_{1},(\forall a=1\,to\,4)\:\:[Q_{a},e_{i}]=\left(e_{i}\right)_{ab}Q_{b}\in L_{1},\,\nonumber \\
and & \,\left\{ Q_{a},Q_{b}\right\} =(N_{i})_{ab}e_{i}\in L_{0}\;(\forall i=1\,to\,3,\forall a,b=1\,to\,4)\},\label{eq:12}
\end{align}
where the $N_{i}\,(\forall i=1\,to\,3)$ must be symmetric, i.e.,
$N_{i}=N_{i}^{T}\,(\forall i=1\,to\,3)$. The $Z_{2}$-graded algebra
$(L$) of quaternions is the direct sum of the two algebras as $L=L_{0}\oplus L_{1}.$
Now, to evaluate the representations of the symmetric $N_{i}$ matrices,
we consider the following generalized Jacobi identity \cite{key-31}:

\begin{align}
[e_{l},\{Q_{a},Q_{b}\}]+\{Q_{b},[e_{l},Q_{a}]\}+\{Q_{a},[Q_{b},e_{l}]\}= & 0.\label{eq:13}
\end{align}
Using the relations of graded Lie algebra defined in eq. (\ref{eq:12}),
we get the following simplified form of eq. (\ref{eq:13}) as

\begin{align}
2(N_{m})_{ab}e_{n}\in_{lmn}+(e_{l})_{ac}(N_{m})_{bc}e_{m}+(e_{l})_{bc}(N_{m})_{ac}e_{m}= & 0,\nonumber \\
e_{l}N_{m}+(e_{l}N_{m})^{T}= & 2\in_{lmn}N{}_{n}\nonumber \\
e_{l}N_{m}-N_{m}e_{l}= & 2\in_{lmn}N{}_{n}\,\:\:\:\:\:\:\:\:(\forall a,b,c=1\,to\,4\:\:\:and\,\,\,\forall l,m,n=1,2,3).\label{eq:14}
\end{align}
The last line of the above eq. is obtained by using the symmetric
property of the $N_{i}$ matrix and noting the transpose properties
of $e_{l}\:\:(\forall l=1,2,3)$ from eq. (\ref{eq:10}). Now, we
define the $N_{i}\,(\forall i=1,2,3)$ matrices as

\begin{align}
N_{i}= & \left[\begin{array}{cccc}
a_{i} & b_{i} & c_{i} & d_{i}\\
w_{i} & f_{i} & g_{i} & h_{i}\\
p_{i} & q_{i} & r_{i} & s_{i}\\
l_{i} & m_{i} & n_{i} & t_{i}
\end{array}\right]\,\,\,\,\,\,\,\,\,\,\,\,\,\,\,\,\,\,\,\,\,(\forall i=1,2,3).\label{eq:15}
\end{align}
Keeping in mind the symmetric conditions imposed on $N_{i}\,'s(\forall i=1,2,3)$
matrices, i.e., $N_{i}=N_{i}^{T}$, we get $b_{i}=w_{i}$, $c_{i}=p_{i},$$q_{i}=g_{i}$,
$l_{i}=d_{i}$, $h_{i}=m_{i}$, and $s_{i}=n_{i}$. Hence, the $N_{1},N_{2},N_{3}$
matrices are defined as:

\begin{align}
N_{1}=\left[\begin{array}{cccc}
a_{1} & b_{1} & c_{1} & d_{1}\\
b_{1} & f_{1} & g_{1} & h_{1}\\
c_{1} & g_{1} & r_{1} & s_{1}\\
d_{1} & h_{1} & s_{1} & t_{1}
\end{array}\right],\:\:N_{2}=\left[\begin{array}{cccc}
a_{2} & b_{2} & c_{2} & d_{2}\\
b_{2} & f_{2} & g_{2} & h_{2}\\
c_{2} & g_{2} & r_{2} & s_{2}\\
d_{2} & h_{2} & s_{2} & t_{2}
\end{array}\right] & ,\:\:N_{3}=\left[\begin{array}{cccc}
a_{3} & b_{3} & c_{3} & d_{3}\\
b_{3} & f_{3} & g_{3} & h_{3}\\
c_{3} & g_{3} & r_{3} & s_{3}\\
d_{3} & h_{3} & s_{3} & t_{3}
\end{array}\right].\label{eq:16}
\end{align}
From eq. (\ref{eq:14}), we have the following relations for $l=m$
$(\forall l=1,2,3)$:

\begin{align}
e_{1}N_{1}-N_{1}e_{1}= & 0\nonumber \\
e_{2}N_{2}-N_{2}e_{2}= & 0\nonumber \\
e_{3}N_{3}-N_{3}e_{3}= & 0.\label{eq:17}
\end{align}
By putting the values of $N_{1}$ from eq. (\ref{eq:16}) and $e_{1}$
from eq.(\ref{eq:10}) into the first relation of the eq.(\ref{eq:14})
above, we get the following conditions for the matrix elements of
$N_{1}$: $2b_{1}=0=2s_{1}$, $f_{1}-a_{1}=0$, $g_{1}-d_{1}=0$,
$h_{1}+c_{1}=0$, and $r_{1}-t_{1}=0$, hence we have the following
form of $N_{1}$:
\begin{align}
N_{1}= & \left[\begin{array}{cccc}
a_{1} & 0 & c_{1} & d_{1}\\
0 & a_{1} & d_{1} & -c_{1}\\
c_{1} & d_{1} & r_{1} & 0\\
d_{1} & -c_{1} & 0 & r_{1}
\end{array}\right].\label{eq:18}
\end{align}
Further, by evaluating the trace of the eq. (\ref{eq:14}), we have:
\begin{align}
Tr(e_{l}N_{m})-Tr(N_{m}e_{l})= & 2\in_{lmn}Tr(N{}_{n})\,\:\:\:\:\:\:\:\:(\forall l,m,n=1,2,3)\nonumber \\
0= & Tr(N_{n})\,\:\:\:\:\:\:\:\:(\forall n=1,2,3).\label{eq:19}
\end{align}
Hence, the trace of $N_{n}'s\:(\forall n=1,2,3)$ matrices must be
equal to zero. Keeping this in mind, we put $a_{1}=r_{1}=0$, and
then the $N_{1}$ matrix has the following form:
\begin{align}
N_{1}= & \left[\begin{array}{cccc}
0 & 0 & c_{1} & d_{1}\\
0 & 0 & d_{1} & -c_{1}\\
c_{1} & d_{1} & 0 & 0\\
d_{1} & -c_{1} & 0 & 0
\end{array}\right].\label{eq:20}
\end{align}
From a similar procedure, using the second relation of eq. (\ref{eq:17})
and the value of $N_{2}$ from eq. (\ref{eq:16}), we evaluate the
following conditions for the $N_{2}$ matrix elements: $2c_{2}=0=2h_{2}$,
$g_{2}+d_{2}=0,$ $r_{2}-a_{2}$=0, $s_{2}-b_{2}=0$, and $t_{2}-f_{2}=0,$
hence we have the following representation for $N_{2}$:
\begin{align}
N_{2} & =\left[\begin{array}{cccc}
a_{2} & b_{2} & 0 & d_{2}\\
b_{2} & f_{2} & -d_{2} & 0\\
0 & -d_{2} & a_{2} & b_{2}\\
d_{2} & 0 & b_{2} & f_{2}
\end{array}\right].\label{eq:21}
\end{align}
Noting the traceless nature of the $N_{2}$ matrix, we further impose
the conditions as $a_{2}=f_{2}=0;$ hence, we have the following form
of the $K_{2}$ matrix:

\begin{align}
N_{2} & =\left[\begin{array}{cccc}
0 & b_{2} & 0 & d_{2}\\
b_{2} & 0 & -d_{2} & 0\\
0 & -d_{2} & 0 & b_{2}\\
d_{2} & 0 & b_{2} & 0
\end{array}\right].\label{eq:22}
\end{align}
Now, from eq. (\ref{eq:14}) for $l=3$ and $m=1$, and using the
values of $N_{1}$, $N_{2}$ from eqs. (\ref{eq:21}) and (\ref{eq:22}),
and $e_{3}$ from eq. (\ref{eq:10}), we may have,
\begin{align}
e_{3}N_{1}-N_{1}e_{3}= & 2N_{2}\nonumber \\
\left[\begin{array}{cccc}
d_{1} & -c_{1} & 0 & 0\\
-c_{1} & -d_{1} & 0 & 0\\
0 & 0 & d_{1} & -c_{1}\\
0 & 0 & -c_{1} & -d_{1}
\end{array}\right]= & \left[\begin{array}{cccc}
0 & b_{2} & 0 & d_{2}\\
b_{2} & 0 & -d_{2} & 0\\
0 & -d_{2} & 0 & b_{2}\\
d_{2} & 0 & b_{2} & 0
\end{array}\right].\label{eq:23}
\end{align}
Hence, we get the conditions: $d_{1}=d_{2}=0$ and $c_{1}=-b_{2}$.
Therefore, we have the following form of $N_{1}$ and $N_{2}$:
\begin{align}
N_{1}= & \left[\begin{array}{cccc}
0 & 0 & c_{1} & 0\\
0 & 0 & 0 & -c_{1}\\
c_{1} & 0 & 0 & 0\\
0 & -c_{1} & 0 & 0
\end{array}\right],\,\,\,\,\,\,\,\,\,N_{2}=\left[\begin{array}{cccc}
0 & -c_{1} & 0 & 0\\
-c_{1} & 0 & 0 & 0\\
0 & 0 & 0 & -c_{1}\\
0 & 0 & -c_{1} & 0
\end{array}\right].\label{eq:24}
\end{align}
By using a similar procedure for the $N_{3}$ matrix and using the
relations evaluated from eq.(\ref{eq:14}) as
\begin{align}
e_{3}N_{3}-N_{3}e_{3}= & 0\nonumber \\
e_{1}N_{2}-N_{2}e_{1}= & 2N_{3}\nonumber \\
e_{3}N_{1}-N_{1}e_{3}= & 2N_{2},\label{eq:25}
\end{align}
we have the following conditions for $N_{3}$ matrix elements : $d_{3}=g_{3}=0,$
$h_{3}=-c_{3}=d_{2}=0,$ $s_{3}=-b_{3}$$=f_{2}=0$, $t_{3}=b_{2}=a_{3}$
$=-f_{3}=-r_{3}=-c_{1}$, $b_{3}=f_{2}=0$, and $a_{2}=-b_{3}=0.$
Hence, we have the following form of the $N_{3}$ matrix: 
\begin{align}
N_{3}= & \left[\begin{array}{cccc}
-c_{1} & 0 & 0 & 0\\
0 & c_{1} & 0 & 0\\
0 & 0 & c_{1} & 0\\
0 & 0 & 0 & -c_{1}
\end{array}\right].\label{eq:26}
\end{align}
By eqs. (\ref{eq:24}) and (\ref{eq:26}), we see that the $c_{1}$
is arbitrary. Now taking the value of $c_{1}=1$, we have the following
matrix representations for $N_{1},N_{2}$, and $N_{3}$:

\begin{align}
N_{1}= & \left[\begin{array}{cccc}
0 & 0 & 1 & 0\\
0 & 0 & 0 & -1\\
1 & 0 & 0 & 0\\
0 & -1 & 0 & 0
\end{array}\right],\,\,N_{2}=-\left[\begin{array}{cccc}
0 & 1 & 0 & 0\\
1 & 0 & 0 & 0\\
0 & 0 & 0 & 1\\
0 & 0 & 1 & 0
\end{array}\right],\,\,N_{3}=\left[\begin{array}{cccc}
-1 & 0 & 0 & 0\\
0 & 1 & 0 & 0\\
0 & 0 & 1 & 0\\
0 & 0 & 0 & -1
\end{array}\right].\label{eq:27}
\end{align}
Together with the matrices in eq.(\ref{eq:10}), they constitute the
graded Lie algebra space of quaternions (eq. (\ref{eq:12})). The
other Jacobi's identity: 

\begin{align}
[Q_{p},\{Q_{q},Q_{r}\}]+[Q_{q},\{Q_{r},Q_{p}\}]+[Q_{r},\{Q_{p},Q_{q}\}]= & 0\:\:\:\:\:(\forall p,q,r=1\,to\,4),\label{eq:28}
\end{align}
can also be shown to be satisfied as well. Using eq. (\ref{eq:12}),
the above identity changes to:
\begin{align}
(N_{j})_{qr}(e_{j})_{ps}+(N_{j})_{rp}(e_{j})_{qs}+(N_{j})_{pq}(e_{j})_{rs}= & 0\:\:\:\:\:(\forall p,q,r,s=1\,to\,4\:\:\:and\:\:\:\forall j=1,2,3).\label{eq:29}
\end{align}

The above equation is again satisfied by the matrix representations
of $e_{j}\,$ and $N_{j}\,$, as can be shown by directly putting
the values of $e_{j}$ and $N_{j}$ matrices from eqs. (\ref{eq:10})
and (\ref{eq:27}) into it. The $N_{r}\,$ matrices defined in eq.
(\ref{eq:27}) are symmetric and unitary as well. Also, they are involutory,
since we have $N_{r}^{2}=I\,(\forall r=1,2,3)$. Now one can see that
the $N_{r}\,$ matrices form the graded Lie algebra representation
in eq. (\ref{eq:12}), hence they will be further abbreviated as graded
partners of quaternion basis units $(e_{i}$) throughout the whole
paper.

It can be seen by simple calculations from eq. (\ref{eq:27}) and
(\ref{eq:10}) that the graded partner matrices $N_{1},N_{2},N_{3}$
satisfy the following multiplication rules: 

\begin{align}
N_{r}N_{s}= & \delta_{rs}-\in_{rst}e_{t}\,\,\,\:\:\:\:\:\:\:(\forall r,s,t=1\,to\,3)\nonumber \\
N_{r}N_{s}+N_{s}N_{r}= & 2\delta_{rs}\,\,\:\:\:\:\:\:\:\:\:\:\:\:\:\:\:\:\:\:\:\:\:\:(\forall r,s=1\,to\,3).\label{eq:30}
\end{align}

Hence, the graded partners $N_{r}$ anti-commute with each other.
Similarly, one can evaluate the multiplication rules between the matrix
representation of quaternion basis units $e_{t}\:(\forall t=1\,to\,3)$
and the symmetric partner matrices $N_{r}\:(\forall r=1\,to\,3)$
as
\begin{align}
N_{r}e_{s}= & \delta_{rs}C+\in_{rst}N_{t}\,\,\,\,\,\,\,\,\,\:\:(\forall r,s,t=1\,to\,3)\nonumber \\
N_{r}e_{s}+e_{s}N_{r}= & 2\delta_{rs}C\:\:\:\:\:\:\:\:\:\:\:\:\,\,\,\,\,\,\,\,\,\:\:\:\:(\forall r,s,t=1\,to\,3).\label{eq:31}
\end{align}
Where the matrix $C$ is evaluated as
\begin{align}
C & =\left[\begin{array}{cccc}
0 & 0 & 0 & -1\\
0 & 0 & -1 & 0\\
0 & 1 & 0 & 0\\
1 & 0 & 0 & 0
\end{array}\right].\label{eq:32}
\end{align}
The matrix $C$ is unitary and has the following properties: $C=-C^{T}=-C^{\dagger}=-C^{-1}$,
with $C^{2}=-I$. The eq. (\ref{eq:31}) also shows that the quaternion
basis units $e_{t}$ and the symmetric partner matrices $N_{r}$ anticommutate
with each other. The multiplication operation of $C$ with the basis
elements $e_{j}\,$ and $K_{r}\,$ maps them into each other, as can
be seen from eqs. (\ref{eq:10}), (\ref{eq:27}), and (\ref{eq:32})
as

\begin{align}
Ce_{j}= & -N_{j}\,\,\,\,\,\,\,\,\,\,\,\,\,\,(\forall j=1\,to\,3)\nonumber \\
CN_{r}= & e_{r}\,\,\,\,\,\,\,\,\,\,\,\,\,\,\,\,\,\,\,(\forall r=1\,to\,3).\label{eq:33}
\end{align}
Now one can evaluate the commutator bracket relations for the quaternion
units $e_{j}\,(\forall j=1\,to\,3)$ and their graded partners $N_{r}\,(\forall r=1\,to\,3)$
with $C$ as

\begin{align}
[e_{i},e_{j}]= & 2\in_{ijk}e_{k}\,\,\,\,\,\,\,\,\,\,\,\,\,\,\,\,\,\,(\forall i,j,k=1\,to\,3)\nonumber \\{}
[N_{r},N_{s}]= & -2\in_{rst}e_{t}\,\,\,\,\,\,\,\,\,\,\,\,\,\,\,(\forall r,s,t=1\,to\,3)\nonumber \\{}
[N_{l},e_{m}]= & 2\in_{lmn}N_{n}\,\,\,\,\,\,\,\,\,\,\,\,\,\,\,(\forall l,m,n=1\,to\,3)\nonumber \\{}
[C,e_{j}]= & 0\,\,\,\,\,\,\,\,\,\,\,\,\,\,\,\,\,\,\,\,\,\,\,\,\,\,\,\,\,\,\,\,\,\,(\forall j=1\,to\,3)\nonumber \\{}
[C,N_{r}]= & 0\,\,\,\,\,\,\,\,\,\,\,\,\,\,\,\,\,\,\,\,\,\,\,\,\,\,\,\,\,\,\,\,\,\,(\forall r=1\,to\,3).\label{eq:34}
\end{align}

Also, from the multiplication rules of eqs. (\ref{eq:30}) and (\ref{eq:31}),
one can see that the quaternion units $e_{j}\,(\forall j=1\,to\,3)$
and their graded partners $N_{r}\,(\forall r=1\,to\,3)$ satisfy the
following Jacobi identities of Lie algebra:
\begin{align}
[e_{l},[e_{m},N_{n}]]+[e_{m},[N_{n},e_{l}]]+[N_{n},[e_{l},e_{m}]]= & 0\:\:\:\:\:\:\:(\forall l,m,n=1\,to\,3)\nonumber \\{}
[e_{l},[N{}_{m},N_{n}]]+[N{}_{m},[N_{n},e_{l}]]+[N_{n},[e_{l},N{}_{m}]]= & 0\:\:\:\:\:\:\:(\forall l,m,n=1\,to\,3)\nonumber \\{}
[N_{l},[N{}_{m},N_{n}]]+[N{}_{m},[N_{n},N_{l}]]+[N_{n},[N_{l},N{}_{m}]]= & 0\:\:\:\:\:\:\:(\forall l,m,n=1\,to\,3).\label{eq:35}
\end{align}

It is to be noted that from the commutation relations of eq. (\ref{eq:34}),
the unit $C$ commutes with both the quaternion $(\mathbb{H})$ basis
units $(e_{i})$ and their graded partners $(N_{r})$; hence, it corresponds
to the Casimir element for the Lie algebra $(T=\{e_{1},e_{2},e_{3},N_{1},N_{2},N_{3}\})$
made by the quaternion $(\mathbb{H})$ basis units $(e_{i})$ and
their graded partners $(N_{r})$. Any state in the linear space of
this Lie algebra $(T=\{e_{1},e_{2},e_{3},N_{1},N_{2},N_{3}\})$ is
mapped by $C$ on to the linear space itself as
\begin{align}
C & |e_{j},N_{l}>=\lambda|-N_{j},e_{l}>\:\:\:\:\:\:\:\:(\forall j,l=1\,to\,3).\label{eq:36}
\end{align}
Also, the Lie algebra $(T)$, which is a non-abelian Lie algebra,
doesn't have any non-zero proper ideals. The Cartan-Killing form for
this Lie algebra is calculated and comes out to be invertible as $B(x,y)=$$Tr(Ad(x)Ad(y))$$=16I_{4}$
\{$\forall x,\,y\in T\}.$

\section{Quaternions $(\mathbb{H})$ and Graded Partner Algebra $(\mathbb{\mathcal{F}}):$}

Now one may define the graded partner algebra $(\mathbb{\mathcal{F}}),$
corresponding to the quaternion algebra $(\mathbb{H}),$ which has
basis units $\{C,N_{1},N_{2},N_{3}\}$. Any graded partner vector
$(F)$ over the field of real numbers is defined in $\mathbb{\mathcal{F}}$
as
\begin{align}
F= & f^{0}N_{0}+f^{1}N_{1}+f^{2}N_{2}+f^{3}N_{3}=f^{0}N_{0}+f^{r}N_{r}\;\;\;\;\;\;\;\;\;\;\;\;(\forall F\subset\mathbb{\mathcal{F}}),\label{eq:37}
\end{align}

where $N_{0}=C$ and $N_{r}\:(\forall r=1\,to\,3)$ are the graded
partner basis units, and $f^{0},$$f^{r}\:(\forall r=1\,to\,3)$ are
the real numbers. Now one can evaluate that the product of any two
graded partner vectors $(F_{1},F_{2})$ in $\mathbb{\mathcal{F}}$
results in a quaternion by using the multiplication rules of eq. (\ref{eq:30})
as 

\begin{align}
F_{1}F_{2}= & (f_{1}^{0}N_{0}+f_{1}^{1}N_{1}+f_{1}^{2}N_{2}+f_{1}^{3}N_{3})(f_{2}^{0}N_{0}+f_{2}^{1}N_{1}+f_{2}^{2}N_{2}+f_{3}^{3}N_{3})\nonumber \\
= & (-f_{1}^{0}f_{2}^{0}+f_{1}^{r}f_{2}^{r})e_{0}+(f_{1}^{0}f_{2}^{r}+f_{1}^{r}f_{2}^{0}-\in_{pqr}f_{1}^{p}f_{2}^{q})e_{r}\:\:\:\label{eq:38}
\end{align}
where $e_{0}$ and $e_{r}\:\:(\forall r=1,2,3)$ are the quaternion
basis units. So, it is clear from eq. (\ref{eq:38}) that the multiplication
operation in the linear space of the graded partner algebra $(\mathbb{\mathcal{F}})$
results in the quaternion linear space $(\mathbb{H})$ as

\begin{align}
:\mathbb{\mathcal{F}}\times\mathbb{\mathcal{F}}\longrightarrow & \mathbb{H}.\label{eq:39}
\end{align}
Also, using the multiplication rules of eq.(\ref{eq:31}), one can
confirm that any multiplication between a quaternion $(Q)$ and graded
partner vector $(F)$ results in a graded partner vector, as
\begin{align}
QF= & (q^{0}e_{0}+q^{1}e_{1}+q^{2}e_{2}+q^{3}e_{3})(f^{0}N_{0}+f^{1}N_{1}+f^{2}N_{2}+f^{3}N_{3})\nonumber \\
= & (q^{0}f^{0}+q^{r}f^{r})N_{0}+(q^{0}f^{r}-q^{r}f^{0}+\in_{str}q^{s}f^{t})N_{r}\:\:\:\:\:\:\:(Q\in\mathbb{H},F\in\mathcal{F}),\label{eq:40}
\end{align}
where $N_{0}$ and $N_{r}\:\:(\forall r=1,2,3)$ are the graded partner
basis units. So, the multiplication operation between any graded partner
vector $(F)$ in $\mathbb{\mathcal{F}}$ and any quaternion element
$(Q)$ results in the space of graded partner algebra $(\mathbb{\mathcal{F}})$
as
\begin{align}
:\mathbb{\mathcal{F}}\times\mathbb{\mathbb{H}}\longrightarrow & \mathbb{\mathbb{\mathcal{F}}}.\label{eq:41}
\end{align}
Keeping in view the positivity and definiteness of the norm of any
vector, we define the norm of a graded partner vector in the space
of graded partner algebra $(\mathcal{F})$ as 
\begin{align}
N(F)=F\widetilde{F}= & (f^{0})^{2}+(f^{1})^{2}+(f^{2})^{2}+(f^{3})^{2}=\left|F\right|\geq0\:\:\:\:\:\:\:\:\:\:\:\:\:\:(\forall F\subset\mathcal{F}).\label{eq:42}
\end{align}
Where $\widetilde{F}$ is the graded partner conjugate of $F$ defined
as
\begin{align}
\widetilde{F}= & -f^{0}N_{0}+f^{1}N_{1}+f^{2}N_{2}+f^{3}N_{3}.\label{eq:43}
\end{align}
The quaternion conjugate of the product of two graded partner vectors
is numerically equal to the product of those graded partner vectors
conjugated separately as
\begin{align}
\overline{F_{1}F_{2}}= & \widetilde{F_{2}}\widetilde{F_{1}},\:\:\:\:\:\:\:\:\:\:\:\:\:\:\:\:\:\:(\forall F_{1},F_{2}\subset\mathcal{F}).\label{eq:44}
\end{align}
Where $\overline{F_{1}F_{2}}$ corresponds to the quaternion conjugate
of $F_{1}F_{2}$ since, according to the eq. (\ref{eq:38}), multiplication
of two graded partner vectors $(F_{1},\,F_{2})$ results in a quaternion,
while $\widetilde{F_{1}}$ and $\widetilde{F_{2}}$ are the graded
partner conjugate vectors of $F_{1}$ and $F_{2}$ defined in eq.
(\ref{eq:43}). The graded partner vectors are not commutative, but
for real parts, they are as
\begin{align}
Re(F_{1}F_{2})= & Re(F_{2}F_{1})=-f_{1}^{0}f_{2}^{0}+f_{1}^{1}f_{2}^{1}+f_{1}^{2}f_{2}^{2}+f_{1}^{3}f_{2}^{3}\:\:\:\:\:\:\:\:\:\:\:\:\:\:\:\:(\forall F_{1},F_{2}\subset\mathcal{F}).\label{eq:45}
\end{align}
The inverse of the multiplication of two graded partners satisfies
the following relation:
\begin{align}
(F_{1}F_{2})^{-1}= & F_{2}^{-1}F_{1}^{-1}\:\:\:\:\:\:\:\:\:\:\:\:\:\:(\forall F_{1},F_{2}\subset\mathcal{F}).\label{eq:46}
\end{align}
Where the inverse of a graded vector $(F^{-1})$ is defined by eq.
(\ref{eq:42}) as
\begin{align}
F^{-1}= & \frac{\widetilde{F}}{\left|F\right|}.\label{eq:47}
\end{align}
Now any graded partner valued $2\times2$ matrix is defined as
\begin{align}
M= & \left[\begin{array}{cc}
m_{1} & m_{2}\\
m_{3} & m_{4}
\end{array}\right],\:\:\:\:\:\:\:\:(\forall m_{1},m_{2},m_{3},m_{4}\in\mathcal{F}).\label{eq:48}
\end{align}
One can evaluate from eq. (\ref{eq:38}) that the multiplication of
two graded partner-valued matrices results in a quaternionic matrix.
It can be seen that due to the anticommutative nature of the multiplication
between graded partner units in eq. (\ref{eq:30}), $(XY)^{T}\neq Y^{T}X^{T}$,
where $X$ and $Y$ are the graded partner valued matrices. Let $X$
and $Y$ be two general graded valued matrices compatible with the
multiplication, then we have:
\begin{align}
(XY)_{ij}^{T}= & (XY)_{ji}\nonumber \\
= & \sum_{k=1}^{n}X_{jk}Y_{ki}\nonumber \\
\neq & \sum_{k=1}^{n}Y_{ki}X_{jk}=\sum_{k=1}^{n}(Y^{T})_{ik}(X^{T})_{kj}=(Y^{T}X^{T})_{ij}\nonumber \\
\therefore\:\:(XY)^{T}\neq & Y^{T}X^{T}.\label{eq:49}
\end{align}

In a similar way to the composition rule of eq. (\ref{eq:40}), we
evaluate that the multiplication of a quaternionic matrix with a graded
partner-valued matrix results in a graded partner-valued matrix. Again,
due to the anticommutative nature of the multiplication between quaternion
basis elements and graded partner units in eq. (\ref{eq:31}), $(XH)^{T}\neq H^{T}X^{T},$
where $X$ is the graded partner valued matrix and $H$ is the quaternion
valued matrix.

The Hermitian conjugate for any $2\times2$ graded partner-valued
matrix $(X)$ is defined as
\begin{align}
M^{\dagger*}=\widetilde{M}^{T}= & \left[\begin{array}{cc}
\widetilde{m}_{1} & \widetilde{m}_{3}\\
\widetilde{m}_{2} & \widetilde{m}_{4}
\end{array}\right],\:\:\:\:\:\:\:\:(\forall m_{1},m_{2},m_{3},m_{4}\in\mathcal{F}).\label{eq:50}
\end{align}
Where $\widetilde{m}_{j}$ is the graded partner conjugate of $m_{j}$
defined in eq.(\ref{eq:43}). 

Also, the quaternionic Hermitian conjugate of the multiplication of
two graded partner matrices is numerically equal to the product of
the Hermitian conjugates of those two graded partner valued matrices
separately:
\begin{align}
\left(XY\right)^{\dagger}= & Y^{\dagger\ast}X^{\dagger\ast}.\label{eq:51}
\end{align}
Where $X$ and $Y$ are two graded partner matrices compatible with
the multiplication and $'\dagger'$ for the quaternionic Hermitian
conjugation operation (because the multiplication $XY$ results in
a quaternionic matrix ) defined in eq. (\ref{eq:8}). One can see
that this matrix multiplication property can be proved easily. Let
the left-hand side of the eq.(\ref{eq:51}) be calculated as 

\begin{align}
(XY)_{ij}^{\dagger}= & (\overline{XY})_{ij}^{T}=\sum_{k=1}^{n}(\overline{X_{jk}Y_{ki}})\nonumber \\
\quad\quad\quad\quad & \quad\quad\quad\quad=\sum_{k=1}^{n}(\widetilde{Y_{ki}}\widetilde{X_{jk}})\nonumber \\
\quad\quad\quad\quad & \quad\quad\quad\quad=\sum_{k=1}^{n}(\widetilde{Y}^{T})_{ik}(\widetilde{X}^{T})_{kj}=(\widetilde{Y}^{T}\widetilde{X}^{T})_{ij}=(Y^{\dagger\ast}X^{\dagger\ast})_{ij}.\label{eq:52}
\end{align}

The second line of the above equation is derived from the multiplication
rule defined in eq.(\ref{eq:44}). Since the multiplication of two
graded partner vectors $X_{jk}$ and $Y_{ki}$ results in a quaternion
element, $\overline{X_{jk}Y_{ki}}$ corresponds to the quaternion
conjugation operation. 

\section{Superspace of Quaternions $(S)$ and Quaternionic Supergroups: }

In the previous two sections we have extended the quaternion Lie algebra
into the $\mathbb{Z}_{2}-graded$ Lie algebra by introducing the graded-partner
matrices. In order to construct a menifest supersymmetric model, the
quaternionic linear vector space has to extended to a superspace.
Basically superspace is the coordinate space of a field theory which
exihibits supersymmetry. Superspace contains all the materials to
define supersymmetric dynamics of the system, same as the usual Minkowski
$(x^{\mu})$ system has do for a non-supersymmetric quantum field
theory. Supersymmetry is the largest extension of the Poincaré symmetry
that contain not only the bosonic degree of freedom but also the fermionic
degree $(\chi_{a},\overline{\chi}_{a})$ of freedom as well. These
fermionic degrees $(\chi_{a},\overline{\chi}_{a})$ are Grassmann
numbers which follow anticommutator relations rather than the commutator
relations. We here use the algebraic properties of the derived graded
partner algebra $(\mathcal{F})$ to construct the superspace of Quaternions
$(\mathbb{H})$.

Now, in order to extend the quaternionic linear vector space $\mathbb{H}^{l}$
over the quaternionic field $\mathbb{H}$ (bosonic sector) \cite{key-20,key-30}
to the quaternionic superspace $(S^{l,m})$, we consider the corresponding
fermionic partners $(\chi_{a})$ in the graded partner vector space
$\mathbb{\mathcal{F}}^{m}$ over the field of graded partner algebra
$\mathbb{\mathcal{F}}$ as,
\begin{align}
\chi_{a}= & \chi_{a}^{0}N_{0}+\chi_{a}^{r}N_{r}\:\:\:\:\:(\forall r=1\,to\,3),\nonumber \\
\{\chi_{a}^{s},\chi_{b}^{t}\}= & 0\:\:\:\:\:\:\:\:(\forall s,t=0,1,2,3).\label{eq:53}
\end{align}

Where $N_{0}=C$ and $N_{r}\:(\forall r=1\,to\,3)$ are the graded
partner basis units of algebra $\mathbb{\mathcal{F}}$. The $l+m$
dimensional quaternionic superspace $S^{l,m}$ is considered to be
described by the coordinates $(q_{1},q_{2},...q_{l},$ $\chi_{1},\chi_{2},...\chi_{m})$
$=s_{A}\in S^{l,m}$ $(A=1,2,...l+m).$ Where $q_{1},q_{2},q_{3},...q_{l}$
are the corresponding quaternionic coordinates in the space of $\mathbb{H}^{l}$
and $\chi_{1},\chi_{2},\chi_{3},...,\chi_{m}$ in the graded partner
vector space $\mathbb{\mathcal{F}}^{m}.$ Therefore, the extended
superspace is created by the two sub-spaces, quaternionic space $\mathbb{H}^{l}$
and symmetric partner space $\mathbb{\mathcal{F}}^{m}$,
\begin{align}
S^{l,m}= & \mathbb{H}^{l}\oplus\mathbb{\mathcal{F}}^{m}.\label{eq:54}
\end{align}

The $S^{l,m}$ is a $\mathbb{Z}_{2}-graded$ superspace, where quaternionic
coordinates ($q_{i})$ construct the even sector $(S^{(0)})$of the
superspace, while the graded partner vectors $(\chi_{j})$ construct
the odd sector $(S^{(1)})$ of the superspace. The multiplication
rules evaluated in eq. (\ref{eq:38}) and (\ref{eq:40}) thus suggest
that
\begin{align}
S^{(i)}\times S^{(j)}= & S^{(i+j)mod\,2},\,\,\,\,\,\,i,j=0,1.\label{eq:55}
\end{align}

Which is the charecteristics of $\mathbb{Z}_{2}-graded$ algebra.
The possible graded matrix representation in the graded linear space
has the following form:
\begin{align}
\varPi(\xi)= & \left(\begin{array}{cc}
C_{1}(\mathbb{H}) & D_{1}(\mathcal{F})\\
D_{2}(\mathbb{\mathcal{F}}) & C_{2}(\mathbb{H})
\end{array}\right),\:\:\:\:\begin{array}{cc}
C_{1}(\mathbb{H})\oplus C_{2}(\mathbb{H}) & \subset Bosonic\;sector\,(\mathbb{H})\\
D_{1}(\mathbb{\mathcal{F}})\oplus D_{2}(\mathbb{\mathcal{F}}) & \subset fermionic\;sector\,(\mathcal{F})
\end{array}.\label{eq:56}
\end{align}

The matrix elements of $\varPi(\xi)$ belong to the $\mathbb{Z}_{2}-graded$
linear superspace $S^{l,m}$. Multiplication between the two graded
matrix matrices in the grded linear space of $S^{l,m}$ retain its
form as the virtue of the algebraic properties of graded partner algebra
$(\mathcal{F})$ and quaternions $(\mathbb{H})$ as
\begin{align}
\varPi_{1}\varPi_{2}= & \left(\begin{array}{cc}
C_{1}^{1} & D_{1}^{1}\\
D_{2}^{1} & C_{2}^{1}
\end{array}\right)\left(\begin{array}{cc}
C_{1}^{2} & D_{1}^{2}\\
D_{2}^{2} & C_{2}^{2}
\end{array}\right)=\left(\begin{array}{ccc}
C_{1}^{1}C_{1}^{2}+D_{1}^{1}D_{2}^{2} & \quad & C_{1}^{1}D_{1}^{2}+D_{1}^{1}C_{2}^{2}\\
\\
D_{2}^{1}C_{1}^{2}+C_{2}^{1}D_{2}^{2} & \quad & D_{2}^{1}D_{1}^{2}+C_{2}^{1}C_{2}^{2}
\end{array}\right)=\left(\begin{array}{cc}
C_{1}^{3} & D_{1}^{3}\\
D_{2}^{3} & C_{2}^{3}
\end{array}\right).\label{eq:57}
\end{align}

where by the multiplication properties defined in eq. (\ref{eq:38})
\& (\ref{eq:40}); $C_{1}^{3}=C_{1}^{1}C_{1}^{2}+D_{1}^{1}D_{2}^{2}$
\& $C_{2}^{3}=D_{2}^{1}D_{1}^{2}+C_{2}^{1}C_{2}^{2}$ $\in\mathbb{H}$,
and $D_{1}^{3}=C_{1}^{1}D_{1}^{2}+D_{1}^{1}C_{2}^{2}$ \& $D_{2}^{3}=D_{2}^{1}C_{1}^{2}+C_{2}^{1}D_{2}^{2}$
$\in\mathbb{\mathcal{F}}$.

Now, we consider the dynamics of the superspace $S^{l,m}$ under the
quaternionic supergroups. The supergroups are the groups of isometries
of the superspace. It has been seen that only two quaternionic supergroups
are possible \cite{key-20}. One is $SL(l,m;\mathbb{H})=SU^{*}(2l,2m)$,
which is not corresponding to any metric-preserving group; the other
is $UU_{a}(l;m;\mathbb{H}),$ which preserves the quaternionic anti-unitary
product. In this case, we have defined the bosonic sector in quaternionic
space and the fermionic sector in graded partner space. The canonical
metric form in the $UU_{a}$ quaternionic supergroup is defined as
$g_{xy}=(I_{l},e_{2}I_{m})$. Now, keeping this in mind, one can extend
the quaternionic anti-unitary product in $\mathbb{H}$ supersymmetrically
to $\mathbb{S}^{l,m}$ space by using eq.(\ref{eq:53}) as

\begin{align}
(S,S')_{UU_{a}}=\bar{S}_{x}g_{xy}S'{}_{y}= & (q,q')_{U}+(\chi,\chi')_{U_{a}}\nonumber \\
=( & q_{k}^{0}q'{}_{k}^{0}+q_{k}^{r}q'{}_{k}^{r}-\chi_{\alpha}^{0}\chi'{}_{\alpha}^{2}+\chi_{\alpha}^{2}\chi'{}_{\alpha}^{0}-\chi_{\alpha}^{3}\chi'{}_{\alpha}^{1}+\chi_{\alpha}^{1}\chi'{}_{\alpha}^{3}\nonumber \\
, & q_{k}^{0}q'{}_{k}^{1}-q_{k}^{1}q'{}_{k}^{0}+q_{k}^{3}q'{}_{k}^{2}-q_{k}^{2}q'{}_{k}^{3}+\chi_{\alpha}^{0}\chi'{}_{\alpha}^{3}+\chi_{\alpha}^{3}\chi'{}_{\alpha}^{0}+\chi_{\alpha}^{1}\chi'{}_{\alpha}^{2}+\chi_{\alpha}^{2}\chi'{}_{\alpha}^{1}\nonumber \\
, & q_{k}^{0}q'{}_{k}^{2}-q_{k}^{2}q'{}_{k}^{0}+q_{k}^{1}q'{}_{k}^{3}-q_{k}^{3}q'{}_{k}^{1}-\chi_{\alpha}^{0}\chi'{}_{\alpha}^{0}+\chi_{\alpha}^{2}\chi'{}_{\alpha}^{2}-\chi_{\alpha}^{1}\chi'{}_{\alpha}^{1}-\chi_{\alpha}^{3}\chi'{}_{\alpha}^{3}\nonumber \\
, & q_{k}^{0}q'{}_{k}^{3}-q_{k}^{3}q'{}_{k}^{0}+q_{k}^{2}q'{}_{k}^{1}-q_{k}^{1}q'{}_{k}^{2}-\chi_{\alpha}^{0}\chi'{}_{\alpha}^{1}-\chi_{\alpha}^{1}\chi'{}_{\alpha}^{0}+\chi_{\alpha}^{3}\chi'{}_{\alpha}^{2}+\chi_{\alpha}^{2}\chi'{}_{\alpha}^{3}),\label{eq:58}
\end{align}
 where $S$ and $S'$ are the elements of superspace $S^{l,m}.$ Here,
the unitary and anti-unitary quaternionic products are defined as:
\begin{align}
(q,q')_{U} & =\bar{q}_{k}q'{}_{k}=(q_{k}^{0}-q_{k}^{1}e_{1}-q_{k}^{2}e_{2}-q_{k}^{3}e_{3})(q'{}_{k}^{0}+e_{1}q'{}_{k}^{1}+e_{2}q'{}_{k}^{2}+e_{3}q'{}_{k}^{3})\nonumber \\
(\chi,\chi')_{U_{a}} & =\chi{}_{\alpha}e_{2}\chi'{}_{\alpha}=(\chi_{\alpha}^{0}C+\chi_{\alpha}^{1}N_{1}+\chi_{\alpha}^{2}N_{2}+\chi_{\alpha}^{3}N_{3})e_{2}(\chi'{}_{\alpha}^{0}C+\chi'{}_{\alpha}^{1}N_{1}+\chi'{}_{\alpha}^{2}N_{2}+\chi'{}_{\alpha}^{3}N_{3}),\label{eq:59}
\end{align}
where $\bar{q}_{k}$ corresponds to the quaternionic conjugation operation.
Comparing this result to Lukierski et al. \cite{key-20}, one can
see that this supersymmetric product is describing the isometries
corresponding to the transformations in the $UU_{a}(l;m;\mathbb{H})$
group, described by the intersections of orthosymplectic supersymmetric
groups as 
\begin{align}
UU_{a}(l;m;\mathbb{H})= & OSp(4l;4m;R)\cap OSp^{(1)}(2l,2l;4m;R)\nonumber \\
\cap & OSp^{(2)}(2l,2l;4m;R)\cap OSp^{(3)}(2l,2l;4m;R).\label{eq:60}
\end{align}

\section{Quaternionic Super conformal algebra in $D=4$ space: }

A conformal transformation of the coordinates is a mapping that leaves
invariant the metric $\zeta_{\mu\nu}$ up to a scale \cite{key-33},
\begin{align}
\zeta'_{\mu\nu}(x')= & \Lambda(x)\zeta_{\mu\nu}.\label{eq:61}
\end{align}
The set of all conformal transformations form a group with the Poincaré
group as a subgroup corresponds to $\Lambda(x)=1$. A conformal group
in $d$ with $p+q=d$ in a metric $\{-1.....,+1....\}$ with signature
$(p,q)$ is isomorphic to the $SO(p+1,q+1)$ group. The conformal
group of $D=4$ dimensional space $\mathbb{R}^{4,0}$ is thus isomorphic
to the $SO(5,1)$ group for Euclidean metric signature. So we have
the $15$ generators for conformal algebra in $D=4$ dimensions, containing
the $6$ generators of rotation $SO(4)$, $4$ generators of translations
$P_{\mu}$ , $4$ generators of conformal accelerations $K_{\mu}$,
and one dilation $D$. 

To construct the $6$ quaternionic generators of $SO(4)$, we first
construct the $2\times2$ dimensional quaternionic valued $\Gamma$-matrices
in Weyl representation by extending the imaginary unit $i$ in the
$\sigma_{2}-$Pauli matrix to triplet quaternionic units $e_{i}$$(\forall i=1,2,\,and\,3)$
as

\begin{align}
\Gamma_{1}= & \left(\begin{array}{cc}
0 & -e_{1}\\
e_{1} & 0
\end{array}\right),\;\Gamma_{2}=\left(\begin{array}{cc}
0 & -e_{2}\\
e_{2} & 0
\end{array}\right),\;\Gamma_{3}=\left(\begin{array}{cc}
0 & -e_{3}\\
e_{3} & 0
\end{array}\right),\Gamma_{0}=\left(\begin{array}{cc}
0 & 1\\
1 & 0
\end{array}\right).\label{eq:62}
\end{align}
These $\Gamma-$matrices satisfy the following relation of Clifford
algebra:

\begin{align}
\Gamma_{\mu}\Gamma_{\nu}+\Gamma_{\nu}\Gamma_{\mu}= & 2\eta_{\mu\nu}I_{2}.\label{eq:63}
\end{align}
Where we have defined $\Gamma_{\mu}=(\Gamma_{k},\Gamma_{0}),$ $(\forall k=1,2,3)$,
and the metric $\eta_{\mu\nu}=\{1,1,1,1\}$. The $\Gamma_{\mu}$ are
in Weyl representation, they satisfy the following properties:

(i) $\Gamma_{0}^{\dagger}=\Gamma_{0}$

(ii) $\Gamma_{k}^{\dagger}=\Gamma_{k}$$\,(\forall k=1,2,3)$

(iii) $\Gamma_{0}\Gamma_{\mu}\Gamma_{0}=\overline{\Gamma}_{\mu}$.

Any arbitrary space-time four vector $(x^{\mu})$ in space may be
associated with these quaternionic Hermitian $\Gamma-$ matrices as
\begin{align}
x^{\mu}\rightarrow X= & x_{\mu}\Gamma_{\mu}=\left(\begin{array}{cc}
0 & x_{0}-x_{1}e_{1}-x_{2}e_{2}-x_{3}e_{3}\\
x_{0}+x_{1}e_{1}+x_{2}e_{2}+x_{3}e_{3} & 0
\end{array}\right).\label{eq:64}
\end{align}
the determinant of which is given \cite{key-14,key-30} as 
\begin{align}
det(X)= & [(x_{0})^{2}+(x_{1})^{2}+(x_{2})^{2}+(x_{3})^{2}]^{2}=(\eta_{\mu\nu}x_{\mu}x_{\nu})^{2}\label{eq:65}
\end{align}
where the metric is $\eta_{\mu\nu}=\{1,1,1,1\}$. The general transformation
from the Hermitian matrix $X$ to another Hermitian matrix $X'$ without
changing the determinant is thus given by
\begin{align}
X\rightarrow X'= & TXT^{\dagger},\,\,\,\,\,\,\,\,\,\,\,(T\in SL(2,\mathbb{H}))\label{eq:66}
\end{align}
where the quaternionic transformation matrix $T$ must have $det(T)=1$,
and $X'=x'_{\mu}\Gamma_{\mu}$. However it is to be noted that it
can't be a general element of $SL(2,\mathbb{H})$. Since the general
line element that remain unchanged of the transformation under \cite{key-30}
$SL(2,\mathbb{H})$ group is a six-dimensional space-time which involve
the matrices $\sigma_{0}=I_{2}$ and $\sigma_{3}$ with the $\Gamma-$matrices
defined in eq. (\ref{eq:62}) as well. $SL(2,\mathbb{H})$ group is
the universal covering group \cite{key-19,key-30} of $SO(5,1)$ i.e.
$SL(2,\mathbb{H})\cong\overline{SO(5,1)}$. Rather under certain conditions
matrices like $T$ form a subgroup of $SL(2,\mathbb{H}),$ we may
call it $Res\,SL(2,\mathbb{H})$ group. So we may have the following
conditions for $T$ matrices transformations:

\begin{align}
T\sigma_{0}T^{\dagger}=\sigma_{0},\quad\quad\quad\quad\quad & T\sigma_{3}T^{\dagger}=\sigma_{3}.\,\,\,\,\,\,\,\,\,\,(\forall T\in Res\,SL(2,\mathbb{H}))\label{eq:67}
\end{align}
Now we define the $T$ matrix as
\begin{align}
T= & \left(\begin{array}{cc}
P & Q\\
R & S
\end{array}\right),\,\,\,\,\,(P,Q,R,S\in\mathbb{H}),\nonumber \\
and & \,\,\,\,det(T)=\left|P\right|^{2}\left|S\right|^{2}+\left|Q\right|^{2}\left|R\right|^{2}-2Re(P\overline{R}S\overline{Q})=1\label{eq:68}
\end{align}
The eq.(\ref{eq:67}) shows that the $T$ matrices must be unitary
hence 
\begin{align}
TT^{\dagger}=\left(\begin{array}{cc}
P & Q\\
R & S
\end{array}\right)\left(\begin{array}{cc}
\overline{P} & \overline{R}\\
\overline{Q} & \overline{S}
\end{array}\right)= & \left(\begin{array}{cc}
1 & 0\\
0 & 1
\end{array}\right).\label{eq:69}
\end{align}
Compairing both sides of the above eq. we get $\left|P\right|^{2}+\left|Q\right|^{2}=\left|R\right|^{2}+\left|S\right|^{2}=1$
and $P\overline{R}+Q\overline{S}=R\overline{P}+S\overline{Q}=0$.
Also by $T^{\dagger}T=I$, we have the conditions $\left|P\right|^{2}+\left|R\right|^{2}=\left|Q\right|^{2}+\left|S\right|^{2}=1$
and $\overline{P}Q+\overline{R}S=\overline{Q}P+\overline{S}R=0$.
By a similar calculations from the second relation of eq. (\ref{eq:67})
we have:
\begin{align}
T\sigma_{3}T^{\dagger} & =\sigma_{3}\nonumber \\
\left(\begin{array}{cc}
P & Q\\
R & S
\end{array}\right)\left(\begin{array}{cc}
1 & 0\\
0 & -1
\end{array}\right)\left(\begin{array}{cc}
\overline{P} & \overline{R}\\
\overline{Q} & \overline{S}
\end{array}\right) & =\left(\begin{array}{cc}
1 & 0\\
0 & -1
\end{array}\right),\label{eq:70}
\end{align}
so, we must have $\left|P\right|^{2}+\left|Q\right|^{2}=1,\left|R\right|^{2}-\left|S\right|^{2}=-1$
and $P\overline{R}-Q\overline{S}=R\overline{P}-S\overline{Q}=0$.
Considering all these conditions together it is concluded that $\left|P\right|^{2}=\left|S\right|^{2}=1$
and $Q=R=0$. Hence the transformation matrices like $T$ which constitute
the group $Res\,SL(2,\mathbb{H})$ must have the following properties:
\begin{align}
T= & \left(\begin{array}{cc}
P & 0\\
0 & S
\end{array}\right),\,\,\,\,\,(P,S\in\mathbb{H}),\,\,\,\,\,\,where\,\,\,\left|P\right|^{2}=\left|S\right|^{2}=1\,\,\,\,\,\nonumber \\
and & \,\,\,\,\,\,TT^{\dagger}=T^{\dagger}T=I\:\:\:\:\,\,\,\,\,\,\,\,\,\,\,\,\,\,\,\,\,\,(\forall T\in Res\,SL(2,\mathbb{H})).\label{eq:71}
\end{align}
 It can be shown that the $Res\,SL(2,\mathbb{H})$ is a closed subgroup
of $SL(2,\mathbb{H})$. Because for any two matrices $T_{1}$, $T_{2}$
$\in Res\,SL(2,\mathbb{H})$, we have,
\begin{align}
T_{1}T_{2}= & \left(\begin{array}{cc}
P_{1} & 0\\
0 & S_{1}
\end{array}\right)\left(\begin{array}{cc}
P_{2} & 0\\
0 & S_{2}
\end{array}\right)=\left(\begin{array}{cc}
P_{1}P_{2} & 0\\
0 & S_{1}S_{2}
\end{array}\right)=\left(\begin{array}{cc}
P_{3} & 0\\
0 & S_{3}
\end{array}\right)\in Res\,SL(2,\mathbb{H}).\label{eq:72}
\end{align}
Where by eq.(\ref{eq:69}) it is easy to show that $(T_{1}T_{2})^{\dagger}(T_{1}T_{2})=T_{2}^{\dagger}T_{1}^{\dagger}T_{1}T_{2}=I$
and $\left|P_{3}\right|^{2}=(\overline{P_{1}P_{2}})(P_{1}P_{2})$
$=\left|P_{1}\right|^{2}\left|P_{2}\right|^{2}=1$ and $\left|S_{3}\right|^{2}=(\overline{S_{1}S_{2}})(S_{1}S_{2})$
$=\left|S_{1}\right|^{2}\left|S_{2}\right|^{2}=1$. Hence, if $T_{1}$
and $T_{2}$ are the elements of $Res\,SL(2,\mathbb{H})$ than $T_{1}T_{2}$
is also an element of this restricted $Res\,SL(2,\mathbb{H})$ group,
which leaves the interval defined in eq. (\ref{eq:65}) invariant.
Also from eq. (\ref{eq:71}) we have $\overline{T}=T^{-1}.$

Since according to the eq. (\ref{eq:71}), the group elements of $Res\,SL(2,\mathbb{H})$
must be unitary, hence $Res\,SL(2,\mathbb{H})$ $\subset$ $U(2,\mathbb{H})\cong Sp(2)$.
Now, it can be shown that the $Res\,SL(2,\mathbb{H})$ group is homomorphic
to the $SO(4)$ group. We may write the $SO(4)$ transformations of
a four vector $x_{\mu}$ in $D=4$ dimensions, which leaves the space
interval defined in eq. (\ref{eq:65}) invariant as
\begin{align}
x'_{\mu}= & \Lambda_{\mu\nu}x_{\nu}\,\,\,\,\,\,\,\,(\forall\mu,\nu=0,1,2,3)\,\,\,\,\,\,\,\,\,\,(\forall\Lambda\in SO(4))\label{eq:73}
\end{align}
Where $\Lambda_{\mu\nu}$ are the matrix components of $SO(4)$ transformation
matrix $(\Lambda)$. Putting the value of $x'_{\mu}$from eq. (\ref{eq:73})
into eq. (\ref{eq:66}) we have;

\begin{align}
\Lambda_{\mu\nu}\Gamma_{\mu}= & T\Gamma_{\nu}T^{\dagger}\,\,\,\,\,\,\,\,\,\,\,(T\in Res\,SL(2,\mathbb{H}))\label{eq:74}
\end{align}

Now using the eq. (\ref{eq:63}), into the eq. above, we get the following
relation between the $SO(4)$ transformation matrix $\Lambda$ and
the matrix $T$ $(\in Res\,SL(2,\mathbb{H})$):
\begin{align}
\Lambda_{\mu\nu}= & \frac{1}{2}Tr[T\Gamma_{\nu}T^{\dagger}\Gamma_{\mu}].\label{eq:75}
\end{align}
 Where the $Tr$ corresponding to the quaternionic trace, i.e. $Tr(Q)=ReTr(Q).$
Hence we may associate a $T$ matrix in $Res\,SL(2,\mathbb{H})$ group
for each and every $SO(4)$ rotation $\Lambda_{\mu\nu}$ in $D=4$
dimensional space. The homomorphism between $SO(4)$ and $Res\,SL(2,\mathbb{H})$
then be described by the relation between the components of $\Lambda_{\mu\nu}$
and $T(\in Res\,SL(2,\mathbb{H}))$. By the simple cyclic properties
of quaternionic trace it is easy to evaluate $\Lambda_{\mu\nu}(T_{1}T_{2})=\Lambda_{\mu\nu}(T_{1})\Lambda_{\mu\nu}(T_{2})$
$\forall\,T_{1},\,T_{2}\in Res\,SL(2,\mathbb{H}).$ Also from eq.
(\ref{eq:71}), we have $\Lambda_{\mu\nu}(T^{-1})=(\Lambda_{\mu\nu}(T))^{-1}$.
The $det(\Lambda)$ is a continuous function of $T(\in Res\,SL(2,\mathbb{H}))$,
since $Res\,SL(2,\mathbb{H})$ is also a continuous group, because
the domain of these variables is simply connected, a discontinuous
jump from $det(\Lambda)=1$ to $det(\Lambda)=-1$ is excluded. By
the consequence for the restriction (\ref{eq:66}) on the values of
$T's$, we have $det(\Lambda)=1$ for all the elements of the $SO(4)$
defined in eq. (\ref{eq:75}). Hence the homomorphism between the
group $SO(4)$ and the group $Res\,SL(2,\mathbb{H})$ has been established
consistently. The vector $X$ defined in eq. (\ref{eq:64}) transform
as a vector under endomorphic transformation in $Res\,SL(2,\mathbb{H})$. 

Similarly, the two-component quaternion spinors acting on $Res\,SL(2,\mathbb{H})$
are defined as

\begin{align}
\varPsi_{\alpha}= & \left(\begin{array}{c}
\phi\\
\xi
\end{array}\right)\:\:(\forall\phi,\,\xi\in\mathbb{H}),\:\:\:\:\:\:\:\:\:\varPsi_{\alpha}^{\dagger}=(\overline{\phi},\,\,\overline{\xi})\label{eq:76}
\end{align}

where $\dagger$ is the quaternionic hermitian conjugate, and $\overline{\phi}$
\& $\overline{\xi}$ are the quaternionic conjugate of $\phi$ \&
$\xi$ respectively. The transformation properties under $Res\,SL(2,\mathbb{H})$
of undotted quaternionic spinor and its conjugate are such as 
\begin{align}
\varPsi'_{\alpha}= & T_{\alpha}^{\:\,\beta}\varPsi_{\beta}\:\:\:\:\:\:\:\:\:\varPsi{}_{\alpha}^{'\dagger}=\varPsi_{\beta}^{\dagger}T_{\:\:\alpha}^{\beta}\quad\quad\quad\quad\quad T\in Res\,SL(2,\mathbb{H}).\label{eq:77}
\end{align}
While the dotted spinors and its conjugate transform as
\begin{align}
\eta^{\overset{.}{\alpha}}= & \left(\begin{array}{c}
\lambda\\
\varsigma
\end{array}\right)\:\:(\forall\lambda,\,\varsigma\in\mathbb{H}),\:\:\:\:\:\:\:\:\:\eta^{\overset{.}{\alpha\dagger}}=(\overline{\lambda},\,\,\overline{\varsigma})\label{eq:78}
\end{align}
where transformation properties under $Res\,SL(2,\mathbb{H})$ are
defined as
\begin{align}
\eta^{\overset{.}{\alpha'}}= & (\overline{T})_{\:\:\overset{.}{\beta}}^{\overset{.}{\alpha}}\eta^{\overset{.}{\beta}},\:\:\:\:\:\:\eta^{\overset{.}{\alpha'}\dagger}=\eta^{\overset{.}{\alpha}\dagger}(\overline{T})_{\overset{.}{\beta}}^{\:\:\overset{.}{\alpha}}.\label{eq:79}
\end{align}
Where we have used the property $T^{-1}=\overline{T}$. The differential
operator is defined as
\begin{align}
\partial=\Gamma_{\mu}\partial_{\mu}= & \left(\begin{array}{cc}
0 & \partial_{0}-e_{1}\partial_{1}-e_{2}\partial_{2}-e_{3}\partial_{3}\\
\partial_{0}+e_{1}\partial_{1}+e_{2}\partial_{2}+e_{3}\partial_{3} & 0
\end{array}\right).\label{eq:80}
\end{align}
 Now one can form the quaternionic $SO(4)$ generators $(\Sigma_{\mu\nu})$
using $\Gamma-$matrices defined in eq. (\ref{eq:62}) as

\begin{align}
\Sigma_{\mu\nu}=\frac{1}{4} & [\Gamma_{\mu}\Gamma_{\nu}-\Gamma_{\nu}\Gamma_{\mu}]\:\:\:\:\:\:\:\:\:\:(\forall\mu,\nu=0\,to\,3).\label{eq:81}
\end{align}
Which satisfy the following $SO(4)$ Lie algebra as 
\begin{align}
[\Sigma_{\mu\nu},\Sigma_{\rho\sigma}]= & -(\eta_{\mu\rho}\Sigma_{\nu\sigma}-\eta_{\mu\sigma}\Sigma_{\nu\rho}-\eta_{\nu\rho}\Sigma_{\mu\sigma}+\eta_{\nu\sigma}\Sigma_{\mu\rho})\;\;\;\;\;\;(\forall\mu,\nu,\rho,\sigma=0\,to\,3).\label{eq:82}
\end{align}
The 4 generators of translation or linear momentum in $D=4$ may then
be written as 

\begin{align}
P_{\mu}= & \frac{1}{2}\left(\begin{array}{cc}
0 & 0\\
e_{\mu} & 0
\end{array}\right)\quad\quad\quad(\forall\mu=0,1,2,3).\label{eq:83}
\end{align}
The 4 generators of conformal accelerations may also be written as
\begin{align}
K_{\mu}= & \frac{1}{2}\left(\begin{array}{cc}
0 & e_{\mu}\\
0 & 0
\end{array}\right)\quad\quad\quad(\forall\mu=0,1,2,3).\label{eq:84}
\end{align}
The generator of dilation is defined as

\begin{align}
D= & \frac{1}{2}\left(\begin{array}{cc}
1 & 0\\
0 & -1
\end{array}\right)\label{eq:85}
\end{align}
One can evaluate that the Lie algebra of the $SO(5,1)$ conformal
group in $D=4$ as
\begin{align}
[p_{\mu},p_{\nu}]= & 0\;\;\;\;\;\;\;\;\;\;\;\;\;\;\;\;\;\;\;\;\;\;\;\;\;\;(\forall\mu,\nu=0\,to\,3)\nonumber \\{}
[K_{\mu},K_{\nu}]= & 0\nonumber \\{}
[\Sigma_{\mu\nu},p_{\rho}]= & -(\eta_{\mu\rho}p_{\nu}-\eta_{\nu\rho}p_{\mu})\:\:\:\:\:\:(\forall\mu,\nu,\rho=0\,to\,3)\nonumber \\{}
[\Sigma_{\mu\nu},\Sigma_{\rho\sigma}]= & -(\eta_{\mu\rho}\Sigma_{\nu\sigma}-\eta_{\mu\sigma}\Sigma_{\nu\rho}-\eta_{\nu\rho}\Sigma_{\mu\sigma}+\eta_{\nu\sigma}\Sigma_{\mu\rho}).\;\;\;\;\;\;(\forall\mu,\nu,\rho,\sigma=0\,to\,3).\nonumber \\{}
[\Sigma_{\mu\nu},K_{\rho}]= & (\eta_{\mu\rho}K_{\nu}-\eta_{\nu\rho}K_{\mu})\:\:\:\:\:\:(\forall\mu,\nu,\rho=0\,to\,3)\nonumber \\{}
[D,P_{\mu}]= & -P_{\mu}\nonumber \\{}
[D,K_{\mu}]= & K_{\mu}\nonumber \\{}
[p_{\mu},K_{\nu}]= & D-\frac{1}{2}\Sigma_{\mu\nu}.\label{eq:86}
\end{align}

Now we introduce the graded matrix representation to form the quaternionic
realization of super Poincaré algebra in $D=4$-dimensional space.
Since Section $5$, we have defined the graded matrix representation
of super algebra in super space $(S)$. We have asserted that the
quaternionic $(\mathbb{H})$ bosonic sector of the super algebra is
contained along the diagonal quadrant, while the fermionic sector
is spanned in the regime of the graded partner algebra $(\mathbb{\mathcal{F}})$
in the off-diagonal quadrant. Hence, keeping in view this form of
graded matrix representation, we introduce the bosonic sector of the
superalgebra as follows:
\begin{align}
M_{\mu\nu}= & \left(\begin{array}{ccc}
\Sigma_{\mu\nu} & \vdots & 0\\
\cdots & \cdots & \cdots\\
0 & \vdots & 0
\end{array}\right),P_{\mu}=\left(\begin{array}{ccc}
p_{\mu} & \vdots & 0\\
\cdots & \cdots & \cdots\\
0 & \vdots & 0
\end{array}\right),\;\;\;\;\;(\forall\mu,\nu=0\,to\,3).\nonumber \\
K_{\mu} & =\left(\begin{array}{ccc}
K_{\mu} & \vdots & 0\\
\cdots & \cdots & \cdots\\
0 & \vdots & 0
\end{array}\right),D=\left(\begin{array}{ccc}
D & \vdots & 0\\
\cdots & \cdots & \cdots\\
0 & \vdots & 0
\end{array}\right)\label{eq:87}
\end{align}

Now to construct the fermionic sector of the $N=1$ super conformal
algebra in $D=4$ dimensions, we introduce the fermionic $4$ super
partners in $D=4$ dimensional space in terms of our graded partner
algebra $(\mathbb{\mathcal{F}})$ as
\begin{align}
Q_{1}(N_{\mu})= & \left(\begin{array}{cccc}
0 & 0 & \vdots & N_{\mu}\\
0 & 0 & \vdots & 0\\
\cdots & \cdots & \cdots & \cdots\\
0 & 0 & \vdots & 0
\end{array}\right),\:\:\:\:\:\:Q_{2}(N_{\mu})=\left(\begin{array}{cccc}
0 & 0 & \vdots & 0\\
0 & 0 & \vdots & N_{\mu}\\
\cdots & \cdots & \cdots & \cdots\\
0 & 0 & \vdots & 0
\end{array}\right)\nonumber \\
S_{1}(N_{\nu})= & \left(\begin{array}{cccc}
0 & 0 & \vdots & 0\\
0 & 0 & \vdots & 0\\
\cdots & \cdots & \cdots & \cdots\\
\widetilde{N}_{\nu} & 0 & \vdots & 0
\end{array}\right),\:\:\:\:\:\:S_{2}(N_{\nu})=\left(\begin{array}{cccc}
0 & 0 & \vdots & 0\\
0 & 0 & \vdots & 0\\
\cdots & \cdots & \cdots & \cdots\\
0 & \widetilde{N}_{\nu} & \vdots & 0
\end{array}\right).\label{eq:88}
\end{align}

Where $N_{\mu}=\{N_{0}=C,N_{1},N_{2},N_{3}\}$ are the basis units
of the graded partner algebra $(\mathbb{\mathcal{F}})$ that follow
the multiplication rules of eqs. (\ref{eq:30}), (\ref{eq:31}), and
(\ref{eq:33}). Hence, keeping in view the graded form of fermionic
super-partners defined in eq.(\ref{eq:88}) and bosonic partners in
eq. (\ref{eq:87}), one can write the complete superalgebra in $D=4$
as
\begin{align}
\{Q_{1}(N_{\mu}),S_{2}(N_{\nu})\}= & \delta_{\mu\nu}N_{0}+\delta_{\mu0}N_{\nu}+\delta_{\nu0}N_{\mu}-\in_{\mu\nu\sigma}N_{\sigma}\;\;(\forall\mu,\nu,\rho,\sigma=0\,to\,3)\nonumber \\
\{Q_{2}(N_{\mu}),S_{1}(N_{\nu})\}= & \delta_{\mu\nu}P_{0}+\delta_{\mu0}P_{\nu}+\delta_{\nu0}P_{\mu}-\in_{\mu\nu\sigma}P_{\sigma}\;\;(\forall\mu,\nu,\rho,\sigma=0\,to\,3)\nonumber \\
\{Q_{1}(N_{\mu}),Q_{2}(N_{\nu})\}= & 0.\:\:\:\:\:\:\:\:\:\:\:\:\:(\forall\mu,\nu=0\,to\,3).\nonumber \\
\{S_{1}(N_{\mu}),S_{2}(N_{\nu})\}= & 0\:\:\:\:\:\:\:\:\:\:\:\:\:\:(\forall\mu,\nu=0\,to\,3)\nonumber \\
\{Q_{1}(N_{\mu}),S_{1}(N_{\nu})\}= & \frac{1}{2}[M_{\mu\nu}+A(\widetilde{N}_{\nu}N_{\mu})+2D(N_{\mu}\widetilde{N}_{\nu})]\nonumber \\
\{Q_{2}(N_{\mu}),S_{2}(N_{\nu})\}= & \frac{1}{2}[M_{\mu\nu}+A(\widetilde{N}_{\nu}N_{\mu})-2D(N_{\mu}\widetilde{N}_{\nu})]\nonumber \\{}
[Q_{1}(N_{\mu}),D]= & -\frac{1}{2}Q_{1}(N_{\mu})\nonumber \\{}
[Q_{2}(N_{\mu}),D]= & \frac{1}{2}Q_{2}(N_{\mu})\nonumber \\{}
[S_{1}(N_{\nu}),D]= & \frac{1}{2}S_{1}(N_{\nu})\nonumber \\{}
[S_{2}(N_{\nu}),D]= & -\frac{1}{2}S_{2}(N_{\nu})\nonumber \\{}
[X_{ij},Q_{k}]= & \delta_{jk}Q_{i}(e_{\mu}N_{\nu})\nonumber \\{}
[X_{ij},S_{k}]= & -\delta_{jk}S_{i}(e_{\mu}\widetilde{N}_{\nu})\label{eq:89}
\end{align}

where $N_{\mu}\widetilde{N}_{\nu}=\delta_{\mu\nu}e_{0}+\delta_{\mu0}e_{\nu}+\delta_{\nu0}e_{\mu}-\in_{\mu\nu\sigma}e_{\sigma}$
and $e_{\mu}\widetilde{N}_{\nu}=-\delta_{\mu\nu}\widetilde{N}_{0}+\delta_{\mu0}\widetilde{N}_{\nu}-\delta_{\nu0}\widetilde{N}_{\mu}+\in_{\mu\nu\sigma}\widetilde{N}_{\sigma}$.
The operator $A=I_{3}$ is the $U(1)$ $R-$symmetry generator, while
$X_{ij}$ are the generators \cite{key-20} of quaternionic $SL(2,\mathbb{H})\cong SU^{*}(4)$
$R-$ symmentry defined as 
\begin{align}
(X_{ij})_{rs}= & e_{\mu}\delta_{is}\delta_{jr},\label{eq:90}
\end{align}
 that satisfy the algebra
\begin{align}
[X_{ij}(e_{\mu}),X_{ab}(e_{\nu})]= & \delta_{ib}X_{aj}(e_{\mu}e_{\nu})-\delta_{ja}X_{il}(e_{\mu}e_{\nu}).\label{eq:91}
\end{align}

\section{Discussion and Conclusions: }

The $\mathbb{Z}_{2}-$ graded algebra of quaternions has been studied
in the matrix representations of real numbers. We have developed the
graded partners $K_{i}$$\,(i=1,2,3)$ by introducing the Lie algebra
of quaternions as the bosonic partner $(L_{0})$ of the graded algebra.
Then, using the Jacobi identities of $\mathbb{Z}_{2}-$ graded algebra,
we have evaluated the matrix representations of graded partners $K_{i}$.
It has been shown that these three matrices, along with the matrix
representations of quaternion units, constitute the $\mathbb{Z}_{2}-$
graded Lie algebra space of quaternions. The multiplication rules
between the graded partner units $K_{i}$ and quaternion basis units
$e_{i}$ have been studied. It has been shown that matrix $C$ commutes
with each and every basis unit $(e_{i}$) of quaternions $(\mathbb{H})$
and graded partner units $(K_{i}$), so it corresponds to the Casimir
element of the Lie algebra $(T=\{e_{1},e_{2},e_{3},K_{1},K_{2},K_{3}\})$,
which is formed by the basis units $e_{i}$ and $K_{i}$. Then we
have defined the graded partner algebra $\mathbb{\mathcal{F}}$ as
comprising the basis units $C$ and $K_{i}$ $(\forall i=1,2,3).$ 

The quaternion space $(\mathbb{H}^{l})$ has been further extended
to the graded superspace $(S^{l,m})$ by considering the bosonic part
of the superspace as quaternionic, while the fermionic part is represented
by the graded partner algebra $(\mathcal{F})$ of quaternions $(\mathbb{H})$.
Then we have defined the graded matrix representation for this superspace,
which has principal diagonal elements as the bosonic part (hence quaternionic
$(\mathbb{H})$) and off-diagonal elements as the fermionic part (made
from $(\mathbb{\mathcal{F}})$). The quaternionic supergroups have
been studied in context with this structure of superalgebra and superspace
by extending the anti-unitary quaternionic product supersymmetrically.

After constructing the superspace as the combined space of quaternion
algebra $(\mathbb{H})$ and graded partner algebra $(\mathcal{F})$,
we have studied the $D=4$ super conformal algebra. Where the bosonic
part of the superalgebra contains the generators of the quaternionic
conformal group, however, the fermionic part of the superalgebra has
been constructed by introducing fermion partners $Q_{a}$ in the field
of graded partner algebra $(\mathbb{\mathcal{F}}).$ Finally, the
super conformal algebra in $D=4$ dimensions has been established.


\begin{thebibliography}{10}
\bibitem[1]{key-1} S. Coleman and J. mandula, Phys. Rev. \textbf{\uline{159}}(1967)\textbf{1251.}

\bibitem[2]{key-2} J. Wess and B. Zumino , Phys. Lett.\textbf{\uline{49B}}(1974)\textbf{52
}\&\textbf{ }Nucl. Phys.\textbf{ }\textbf{\uline{B70}} (1974)\textbf{39.} 

\bibitem[3]{key-3} R. Haag J. T. Lopuzanski and M. F. Sohnius, Nucl.
Phys. \textbf{B}\textbf{\uline{ 88}},(1975)\textbf{257}.

\bibitem[4]{key-4} M.F. Sohnius, Physics Reports (Review Section
of Physics Latters) \textbf{\uline{128}} Nos. \textbf{2\& 3}(1985)
\textbf{39-204}, North-Holland, Amsterdam. 

\bibitem[5]{key-5} L. Brink, J.H. Schwarz, J. Scherk, Nuclear Physics
\textbf{B}\textbf{\uline{121}}\uline{(}1977) \textbf{77-92}.

\bibitem[6]{key-6} W.R. Hamilton, \textbf{``Elements of Quaternions}'',
Vol. \textbf{I \& II} Chelsea Publishing, New York (1969).

\bibitem[7]{key-7} G. M. Dixon, \textbf{``Division Algebras: Octonions,
Quaternions, Complex Numbers and the Algebraic Design of Physics''},
Springer- Science +Buisness Media, B.V.(1994). 

\bibitem[8]{key-8} J. H. Conway, D.A. Smith, \textbf{``On quaternions
and Octonions their geometry, airthmatic and symmetry''}, A. K. peters
(2003).

\bibitem[9]{key-9} P.G.Tait, \textbf{\textquotedbl An elementary
Treatise on Quaternions\textquotedbl}, Oxford Univ. Press (1875). 

\bibitem[10]{key-10} Feza Gursey, Chia-Hsiung Tze, \textquotedblleft{}
\textbf{On the Role of Division, Jordan and Related Algebras in Particle
Physics}\textquotedblright , World Scientific Publishing Company (1996).

\bibitem[11]{key-11} Bhupendra C. S. Chauhan and O. P. S. Negi, Fundamental
J. Math. Physics, \textbf{\uline{1(1)}}(2011)\textbf{41-52.} 

\bibitem[12]{key-12} B. C. Chanyal, Int. J. Mod. Phys. A \textbf{\uline{34(31)}}
(2019)\textbf{1950202.}

\bibitem[13]{key-13} B. C. Chanyal and Sandhya Karnatak, Int. J.
Geo. Methods Mod Phys \textbf{\uline{17(02)}}(2020)\textbf{2050018}.

\bibitem[14]{key-14} K. Morita, Progress of Theoretical Physics,
\textbf{\uline{117(3)}}(2007)\textbf{501\textendash 532}.

\bibitem[15]{key-15} J. C. Baez, Bull. Amer. Math. Soci., \textbf{\uline{39}}(2002)\textbf{.
145.}

\bibitem[16]{key-16} J. C. Baez and J. Huerta, arXiv:\textbf{hep-th/0909.0551v2
(2009). }

\bibitem[17]{key-17} Susumo Okubo, \textbf{``Introduction to Octonion
and Other Non-Associative Algebras in Physics'', }Cambridge University
Press\textbf{ }(2009)\textbf{.}

\bibitem[18]{key-18} Seema Rawat and O.P.S. Negi, Int. J. Theor.
Phys. \textbf{\uline{48}}(2009) \textbf{305.}

\bibitem[19]{key-19} T. Kugo and Paul Townsend, Nuclear Physics \textbf{B}\textbf{\uline{221}}
(1983) \textbf{357-380}. 

\bibitem[20]{key-20} Jerzy Lukierski and Anatol Nowicki, Annals of
Physics \textbf{\uline{166}} (1986)\textbf{164-188}.

\bibitem[21]{key-21} Jerzy Lukierski and Francesco Toppan arXiv:
\textbf{hep-th/0203149v1 (2002).}

\bibitem[22]{key-22} J. M. Evans, Nuclear Physics \textbf{B}\textbf{\uline{298}}(1988)
\textbf{92-108.}

\bibitem[23]{key-23} A. Anastasiou, L. Borsten, M. J. Duff, L. J.
Hughes, S. Nagy, JHEP \textbf{\uline{1408}}(2014)\textbf{080}.

\bibitem[24]{key-24} Feza G$\overset{..}{u}$rsey, Modern Physics
Letters A, \textbf{\uline{2,12}}(1987) \textbf{967}. 

\bibitem[25]{key-25} M. G$\overset{..}{u}$naydin, IL NUOVO CIMENTO,
\textbf{\uline{29A}}(1975)\textbf{467}. 

\bibitem[26]{key-26} A. Reit D$\overset{..}{u}$ndarer, Feza G$\overset{..}{u}$rsey,
J. Math. Phys. \textbf{\uline{32}}(1991)\textbf{1176}.

\bibitem[27]{key-27} Corinne A. Manogue, Jörg Schray, J.Math.Phys.
\textbf{\uline{34}}(1993)\textbf{3746.} 

\bibitem[28]{key-28} K. W. Chung, A. Sudbery, Phys. Lett. B, \textbf{\uline{198}}(1987)\textbf{161}.

\bibitem[29]{key-29} A. Sudbery, J. Phys. A: Math. Gen. \textbf{\uline{17}}(1984)\textbf{939}. 

\bibitem[30]{key-30} Bhupendra C. S. Chauhan, Pawan Kumar Joshi and
O. P. S. Negi , Int. J. Mod. Phys. A \textbf{\uline{34(01)}}(2019)\textbf{1950006.} 

\bibitem[31]{key-31} H. J. W. M$\overset{..}{u}$ller Kirstein and
A. Wiedemann, \textbf{``Supersymmetry''}, Word Scientific (1987).

\bibitem[32]{key-32} Robert Gilmore, \textquotedblleft \textbf{Lie
Groups, Physics and Geometry: An Introduction for Physicists, Engineers
and Chemists}\textquotedblright{} , Cambridge University Press (1996). 

\bibitem[33]{key-33} Philippe D. Francesco, Pierre Mathieu, David
Sénéchal, ``\textbf{Conformal Field Theory}'', Springer (1997).
\end{thebibliography}
\end{document}